\documentstyle[preprint,aps,version2]{revtex}
\tightenlines
\begin{document}
\preprint{RBI-TP-1/93}
\begin{title}
IS THERE ANOTHER SCALE IN THE FRACTIONAL \\
QUANTUM HALL EFFECT\@?
\end{title}
\author{ R. BRAKO and \v Z. CRLJEN}
\begin{instit}
Rudjer Bo\v{s}kovi\'{c} Institute, P.O.~Box 1016, 41001 Zagreb, Croatia
\end{instit}
\vspace{2cm}
\begin{abstract}

We reanalyse theoretical considerations and experimental data,
in an attempt to decide
whether there is another scale in the
fractional quantum Hall effect problem, in addition to
the magnetic scale defined by the magnetic length $a_c$ or
the cyclotron energy $\hbar \omega_c$.
We then discuss possible
implications of a new scale on the formulation
of a theoretical model of the fractional quantum Hall effect.
\end{abstract}
\newpage
Since its discovery, the fractional quantum Hall effect (FQHE)$^{\ref{tsui})}$
has attracted considerable attention. The effect occurs in two-dimensional
electron systems (realized at certain
semiconductor interfaces), with applied magnetic fields so strong
that only a fraction of the lowest Landau level is occupied. The
questions to be answered are the same as for the integer
quantum Hall effect (IQHE), namely, why the Hall conductance is
very accurately quantized in multiples (IQHE) or rational fractions
(FQHE) of the fundamental quantity $e^2/h$,
in spite of the irregularities present in
real samples, and why the conductance plateaus occur over a
finite interval of the magnetic field $B$.
The answers are now rather well understood.
An additional and most intriguing question in the FQHE is why
the effect exists at all.
In the IQHE, there is a large energy gap $\hbar \omega_c$ between
Landau levels even in the idealized picture of non-interacting electrons
and no irregularities, which is a good starting point for
understanding the question
why the states with an integer number of  Landau levels
filled are particularly stable. There is no similar simple
argument in the FQHE. Obviously, the electron-electron interaction
must be taken into account if any non-trivial property is to
be obtained, and even then it is not clear at first sight why
certain filling fractions (notably $\nu = 1/m$, where
$m=1, 3, 5 \ldots$) are energetically favoured.

Much of the present understanding of the FQHE is based on
strongly correlated many-body wavefunctions proposed by
Laughlin$^{\ref{prangebook})}$.
Their simplest form is
\begin{equation}
\psi_m = \prod_{j<k}^J (z_i - z_k)^m \exp \left\{ -  \frac{1}{4}
\sum_l^J \left| z_l \right| ^2 \right\},
\label{func}
\end{equation}
where $z_i = x_i + i y_i$ is the coordinate of the $i$-th
electron and the magnetic length $a_c$ has been
set equal to one. These wavefunctions describe a circular droplet of $J$
electrons with a constant density, corresponding to a filling
fraction $\nu = 1/m$. It can be easily seen that they consist
entirely of one-electron states from the lowest Landau level, $n=0$.
This approximation has been
justified by the fact that at large fields the
energy gap between Landau levels is much larger than the
effects we are looking for. Calculations$^{\ref{prangebook})}$
have indicated that
Laughlin's wavefunctions are indeed more stable than other states
at the same filling factor, and that there is an energy gap
which makes them incompressible. (Various modifications have
been proposed in order to explain the FQHE
at other filling factors, or to treat other geometries.
We shall not consider them here.) Laughlin's wavefunctions
are the eigenfunctions of the short-range interaction of the
type$^{\ref{trugman})}$
\begin{equation}
V(r) = \sum_m V_m \, \nabla^{2m} \, \delta^2({\bf r}). \label{pot}
\end{equation}
More generally,
it seems that in order to have a finite energy gap to excited states,
the interaction potential must be singular enough.
Numerical diagonalisation of small systems shows that the true ground
state is close to the corresponding Laughlin's function for any
such potential, and, in particular, for the Coulomb potential.
These calculations are, however, limited to the
lowest Landau level and a small number of particles, so that the
generality of their results must be regarded with some caution.
Anyway, it is widely accepted that a theory which includes states only
from the lowest Landau level and assumes a strong repulsive
interaction describes the FQHE correctly. Eq.~(\ref{func}) is
believed to describe the qualitative properties correctly, such as the
existence of the energy gap. In the following we point out that
there are certain difficulties with this approach.
It has recently been pointed out$^{\ref{brako},\ref{zhang})}$ that it is
inconsistent to assume at the same time that the potential is
extremely short-ranged and that only the states from the lowest
Landau level are involved. We shall first list
some general properties of short-range potentials in quantum
mechanics, both with and without the magnetic field,
then re-examine the FQHE in the light of these and,
finally, confront the experimental data with the theory.

An overview of
the properties of $\delta$-function potentials in quantum mechanics
has recently been given by Gosdzinsky and Tarrach$^{\ref{gosdzinsky})}$.
Using the regularisation procedure, it can be shown
that attractive $\delta$-potentials in two and three spatial
dimensions can give an interacting theory if the coupling constant
(potential strength) approaches zero in an appropriate way
as the range of the potential is made to vanish. In two
dimensions, there is a bound state, in three, the potential
only scatters. On the other hand, repulsive potentials in
more than one dimension and attractive in more than three
dimensions do not lead to interaction.

In two dimensions, a magnetic field perpendicular to the
plane closes the electron orbits, and the energy spectrum becomes
discrete. The problem of a potential centre thus appears more
similar to the perturbation calculation for bound states than
to that for scattering.
In the case of a $\delta$-function potential,
one can proceed by taking a finite number
$N$ of Landau levels instead of using a regulator$^{\ref{brako})}$.
A potential centre at $z=0$ leads to an energy
shift of $m=0$ states from the unperturbed values
$\hbar \omega_c \left(n + 1/2 \right)$. If the
potential contains derivatives, such as the higher terms in
Eq.~(\ref{pot}), the states with the corresponding $m$ are
affected. In all cases, the shifts for low $n$'s vanish as
the number of Landau levels included goes to infinity.
The only exception is the lowest state when the potential is attractive,
which has a large negative energy shift, and can be made to converge
to a bound state of a finite energy
if the strength of the potential approaches zero in a particular
way when $N \rightarrow \infty$. These results are analogous
to those in the case without the magnetic field, if the phrase
`a scattering phase shift' is replaced by `an energy shift'. In
particular, there is no residual interaction when the potential is
repulsive and $N \rightarrow \infty$.

In order to apply these results to real systems, we assume
that the cutoff energy defined by $D=\hbar \omega_c N$ depends upon the
material, but not upon the strength of the magnetic field.
In the following
we divide all energies by $D$, making them proportional to
`real' units, while the magnetic field becomes proportional to $1/N$.
The most interesting quantity is the shift of the
lowest Landau level in the limit of a very strong repulsive
potential$^{\ref{brako})}$:
\begin{equation}
\Delta_m \approx
\left[ N \sum^N_{j=1}{1 \over j}\left(\begin{array}{c} j+|m| \\ |m|
\end{array} \right) \right]^{-1}.
\label{shift}
\end{equation}
This in a sense measures the strength of the effective potential
felt by the electrons in the lowest Landau level.
In the weak-field limit, the leading behaviour is
$N^{-m-1}$, i.e. $B^{m+1}$, but higher terms become quickly
important. For example:
\begin{displaymath}
\Delta_0 \approx
[N(\gamma + \ln N)]^{-1},
\end{displaymath}
\begin{equation}
\Delta_1 \approx
[N(N+\gamma + \ln N)]^{-1},
\label{shiftapprox}
\end{equation}
\begin{displaymath}
\Delta_2 \approx
\left[ N \left( {{N^2} \over 4} + {7 \over 4} N +\gamma + \ln N
\right) \right]^{-1},
\end{displaymath}
where $\gamma$ is Euler's constant.
We have formulated our results in terms of an energy cutoff. By
virtue of quantum-mechanical uncertainty relations, the latter
is related to a spatial cutoff, which is more transparent physically.
A large but finite energy cutoff is equivalent to a potential of
a very short but finite range.

For long-range potentials, there is little connection, if any,
between the problem of scattering on a fixed potential centre
and the many-body problem with the interaction potential of the
same form. This is not so in our case.
In Refs.~\ref{brako} and \ref{zhang} it has been argued that
in two-dimensional systems with an applied magnetic field
the results
concerning the effective strength of short-range interactions
remain valid in the many-particle case, too.
The most interesting consequence is
that the effective interaction disappears if all Landau levels are
taken into account. Including all Landau levels
amounts to saying that no deviation from an ideal
two-dimensionality is observed at any arbitrarily high energy.
This vanishing of the effective interaction
is in sharp contrast to what is
obtained when Laughlin's wavefunctions (which only contain states
from the lowest Landau level) and a short-range interaction are
used to describe the FQHE. A way out of this apparent contradiction
is to argue that the prescription `Laughlin's wavefunctions plus a
$\delta$-function interaction' must not be taken literally, in spite
of its appealing simplicity,  and that the true finite-range
interaction would give similar results,
while being much less sensitive upon the
inclusion of higher Landau levels. Thus
the most likely chain of arguments
is that the starting model for the FQHE problem should be
`an ideal two-dimensional electron gas with an (approximately)
Coulomb interaction'. This model maps with great accuracy onto
`lowest Landau level states with a short-range interaction', which
is in turn diagonalised by
Laughlin's wavefunctions at filling factors $\nu = 1/m$.

However, difficulties become evident even with this interpretation
under further analysis.
A crucial property of the model of `an ideal two-dimensional
electron gas with a Coulomb interaction' is that
it has {\it only one scale}, the magnetic scale defined by
$a_c$ or $\hbar \omega_c$.
The repulsive Coulomb interaction defines no scale of its own,
which can be seen from the fact that the Rutherford
scattering is classical.
The average distance between electrons is also proportional
to the magnetic length, because the FQHE occurs at constant
values of the filling factor. A consequence of the single scale is
that the strength of the electron-electron interaction
(and hence of the magnitude of the FQHE gap) must vary as the
average Coulomb repulsion, i.e. as $e^2/a_c \sim B^{1/2}$, when
the magnetic field $B$ is varied. This argument is general,
and does not depend upon the use of Laughlin's wavefunctions or
any other approximate approach.
One is thus led to the conclusion that the experimental gap
{\it must} scale with $B^{1/2}$, unless some physics beyond the model
of  `an ideal two-dimensional electron gas with a Coulomb interaction'
is relevant. Before making further theoretical considerations, we look
at experimental results.

The energy gap $\Delta$ has been determined experimentally by
measuring the thermal activation behaviour of the diagonal
resistivity at temperatures below 1K, with the strength of the magnetic
field corresponding to the centres of the FQHE plateaus%
$^{\ref{clark88}-\ref{wakabayashi})}$.
In Fig.~\ref{fig1} we show a log-log plot of the magnitudes of the
gap vs. the magnetic field. A theory which
takes into account only the lowest Landau level implies the
electron-hole symmetry within the Landau level, and hence the
equivalence of the FQHE states at filling factors $1/3$ and $2/3$,
$1/5$ and $4/5$, $2/5$ and $3/5$, etc.
While in the following we argue that the restriction
to the lowest Landau level does
not correctly give the  absolute values of the energy gaps
and their scaling with the magnetic field, we expect
that {\it at any value of the field} the equivalence still holds to
a good approximation.
This means that the gap of the $1/3$ state
is very similar to what the gap of the $2/3$ state at the same
field would be (which has not been verified experimentally,
because there is no way to vary the concentration of the electrons
in the layer by such a large factor).
The reason for this similarity is that the electron-hole
symmetry argument gives the same value of the gap for both states,
and that the reduction by the higher Landau level is rather similar.
We have connected the experimental points referring to
the same sample and to the equivalent filling factors. Note that
the states at $1/5$ and at $2/5$ are not equivalent, and, indeed,
the experimental values of the energy gaps are widely apart.
In Ref.~\ref{mallett} the data were interpreted by a $B^{1/2}$
dependence, but with a constant negative offset due to the
disorder present in the sample. According to this interpretation,
a (sample-dependent) threshold magnetic field should exist at which
the FQHE gap becomes zero, and below which there is no FQHE. The
agreement was poor. Our plot shows no sign of a threshold.
The points lie on straight lines, with slopes depending upon the
denominator of $\nu$. This corresponds to a power dependence upon
$B$, with powers clearly larger than $1/2$, which is
reminiscent of our result for a single potential
centre (\ref{shiftapprox}). It seems to us that
this dependence is genuine, and not a consequence of the
disorder, because the behaviour is more universal at
low fields than at higher fields, where there is a large scatter
of data. Thus we reject the interpretation
in terms of a  $B^{1/2}$ dependence shifted by disorder, and conclude
that there must be another scale in the problem that
depends upon the properties of the medium and is independent of
$B$. In the limit $B \rightarrow 0$, the magnetic length
becomes large, while other medium-dependent scales stay constant
(the width of the electron layer, the small-distance unscreened
portion of the Coulomb interaction, etc.; the new scale must result
from some of these). In other words, this is the
limit of the ideal two-dimensionality.
The fact that in this limit
no residual $B^{1/2}$ dependence of the gaps is observed leads to
the conclusion that in a strictly two-dimensional electron gas a
bare Coulomb interaction produces no FQHE at all.

This very surprising statement runs contrary to
the general belief, and we must consider whether it is in
contradiction with any firmly established theoretical results.
We think that this is not the case. The calculations which give
a definite numerical prediction for the FQHE gaps are based
either on numerical diagonalisation within the
lowest Landau level or upon Laughlin's wave functions,
which also imply the restriction to the lowest Landau level.
This is an artificial restriction of the Hilbert space of
physical states. The conclusion reached in the preceding
paragraph suggests that the gaps would vanish if higher
Landau levels were included, which is at present impossible to
verify numerically. (The inclusion of one or several higher
Landau levels cannot give conclusive results.) There are
other calculations,
e.g. based on the Landau-Ginsburg approach, which do not make
the restriction to the lowest Landau level. To our knowledge,
these calculations have proved that the symmetry of the
ground-state wave function at rational filling factors
coincides with that of the corresponding Laughlin's functions,
but they have not been successful in calculating the FQHE gap,
and even the argument that the state is incompressible (i.e.
that the gap is finite) depends on unproved additional
assumptions.

We therefore suggest that, in real systems, the FQHE depends
upon the existence of another scale.
Judged from the results for a single
potential centre, the energy scale is large compared with
the characteristic energy $\hbar \omega_c$ of the problem, i.e.
the spatial scale is small compared with $a_c$. The origin of
this scale must lie beyond the usual assumptions, which are:
(a) strict two-dimensionality, (b) Coulomb interaction, (c)
translational invariance, i.e. no impurities.
A possible candidate is the screening of the
Coulomb interaction in real semiconductor devices, which
modifies the assumption (b).
The effective interaction varies from the bare Coulomb
form $r^{-1}$ at distances smaller than, say, the interatomic
separation, to the screened one, $(r\epsilon)^{-1}$, where
$\epsilon$ is the dielectric constant of the surrounding
medium at large distances.
Another possibility is that the new scale is associated with
relaxing of assumption (a), i.e. that
the motion of the electron in the third dimension
becomes important. The characteristic length is the width of the
potential well which binds the electron gas to the interface,
and the corresponding characteristic energy is that
of the first excited state of the perpendicular motion.
Taking into account this degree of freedom invalidates the
assumption of the perfect two-dimensionality, but only on an
energy scale which is much larger than the characteristic
energies of the system, i.e. the FQHE gaps.
This modification of the model has deep consequences,
because some properties that depend upon strict
two-dimensionality, such as the possibility of performing the
transformation of electrons into `anyons', particles with
arbitrary quantum statistics, become only
approximative. The mechanism which allows the high-energy
properties to become relevant can be visualised in the
following way. Quantum fluctuations can bring two
electrons so close one to the other that the energy of
the repulsive Coulomb interaction equals that of the
first excited state of the  perpendicular motion. Then
the electrons can exchange their positions by one passing
`above' the other, i.e. in the `high-energy physics' of
the system, the trajectories by which the electrons exchange
their positions avoiding each other on one or the other side
are no longer topologically inequivalent, in contrast to the
requirement of
the theories which perform a transformation to anyons.
The Coulomb interaction is essential, but only its
short-range part is selected by this mechanism.

In order to clarify the origin of the proposed new scale, an
improved theoretical treatment of various models is necessary.
Should such a treatment show that the
standard model of two-dimensional electrons with a Coulomb
interaction would give the energy gaps in the FQHE even when the higher
Landau levels would be taken into account, the unusual low-field
dependence of the gaps would remain to be explained. We do not
think that this interpretation is probable, although there
is still the possibility
that a new scale is induced by  the disorder in the sample, which
causes a profound change of the dependence of the gaps
upon the magnetic field. We consider it more probable that the new
scale is due to a modification of the other two assumptions of the
original model. This could be either the dielectric screening of
the electron-electron interaction, which should be possible to
prove theoretically, or, if it turns out that a two-dimensional
theory is not sufficient, the fact that the electrons in real
samples are bound to a plane only by a finite potential well.
This latter possibility seems the most appealing.

To conclude, the analysis of experimental data makes us
believe that there is another scale in the FQHE problem.
A comparison with theoretical calculations on the effect
of short-range potentials suggests that the scale corresponds to
an energy large compared with, say, the FQHE gap, or to a
length small compared with the magnetic length. The most likely
mechanism to generate this scale is the three-dimensionality
of the real system, which becomes evident
at high energies. The low-energy consequence
of this is the opening of finite FQHE gaps.
At present, we are not able to propose a full theoretical
treatment of the problem.

\newpage
\begin{center} References \end{center}

\vspace{10mm}

\begin{list}%
{\arabic{enumi})}{\usecounter{enumi}
    \setlength{\rightmargin}{\leftmargin}}

\item \label{tsui} D.C.~Tsui, H.L.~Stormer and A.C.~Gossard,
Phys.\ Rev.\ Lett.\ {\bf 48} (1982) 1559;
\item \label{prangebook} {\it The Quantum Hall Effect}, edited by
R.E.~Prange and S.M.~Girvin (Springer-Verlag, New York, 1987);
\item \label{trugman} S.A.~Trugman and S.~Kivelson,
Phys.\ Rev.\ B {\bf 31} (1985) 5280;
\item \label{brako} R.~Brako and \v Z.~Crljen, Phys.\ Rev.\ B
{\bf 47} (1993) 13568;
\item \label{zhang} Fu~Chun~Zhang, M.~Ma, YuDong~Zhu and J.K.~Jain,
Phys.\ Rev.\ B {\bf 46} (1992) 2632;
\item \label{gosdzinsky} P.~Gosdzinsky and R.~Tarrach, Am.\ J.\ Phys.
{\bf 59} (1991) 70;
\item \label{clark88} R.G.~Clark, R.J.~Mallett, S.R.~Haynes,
J.J.~Harris and  C.T.~Foxon, Phys.\ Rev.\ Lett.\ {\bf 60} (1988) 1747;
\item \label{mallett} R.J.~Mallett, R.G.~Clark, R.J.~Nicholas,
R.~Willett, J.J.~Harris
and C.T.\ Foxon,  Phys.\ Rev.\ B {\bf 38} (1988) 2200;
\item \label{ebert} G.~Ebert, K.~von~Klitzing, J.C.~Maan,
G.~Remenyi, C.~Probst,
G.~Wei\-mann and W.~Schlapp, J.\ Phys.\ C {\bf 17} (1984) L775;
\item \label{clark86} R.G.~Clark, R.J.~Nicholas, A.~Usher,
C.T.~Foxon and J.J.~Harris, Surface Sci.\ {\bf 170} (1986) 141;
\item \label{wakabayashi} J.~Wakabayashi, S.~Kawaji, J.~Yoshino
and H.~Sakaki, J.\ Phys.\ Soc.\ Japan {\bf 55} (1986) 1319.

\end{list}

\newpage
\figure{ Experimental values of the FQHE gap as a function of the
magnetic field, for several filling factors $p/q$. Diamonds:
$q=3$, and, in order from left to right, $p=5$, 4, 2; full
squares: $q=5$, $p=8$, 7, 3, 2; crosses: $q=7$, $p=10$, 9, 4, 3;
triangles: $q=9$, $p=5$, 4 (all from Ref.~\ref{clark88}); empty
squares: $q=5$, $p=1$ for both points (Ref.~\ref{mallett}).
\label{fig1}}

\end{document}